\documentclass[a4paper,11pt]{article}
\pdfoutput=1 

\usepackage{jcappub} 

\usepackage[T1]{fontenc} 
\usepackage{ulem} 

\usepackage{comment}

\title{Constraining small-scale primordial magnetic fields from the abundance of primordial black holes}

\author[a,b]{Ashu Kushwaha,}
\author[c]{and Teruaki Suyama}

\affiliation[a]{Department of Physics, Indian Institute of Science C. V. Raman Road, Bangalore 560012, India}
\affiliation[b]{Department of Physics, Indian Institute of
  Technology Bombay, Mumbai 400076, India}
\affiliation[c]{Department of Physics, Tokyo Institute of Technology, 2-12-1 Ookayama, Meguro-ku,
Tokyo 152-8551, Japan}

\emailAdd{ashuk@iisc.ac.in, suyama@phys.titech.ac.jp}

\abstract{The presence of magnetic fields in the early universe affects the cosmological processes, leading to the distinct signature, which allows constraining their properties and the genesis mechanisms. In this study, we revisit the method to constrain the amplitude of the magnetic fields on small scales in the radiation-dominated era from the abundance of primordial black holes. 
Constraints in the previous work were based on the fact that the density perturbations sourced by stronger magnetic fields become large enough to gravitationally collapse to form PBHs. However, we demonstrate that this picture is incomplete because 
magnetic fields also increase the threshold value of the density contrast 
required for PBH formation. 
The increase in threshold density contrast is more pronounced on smaller scales,
and in extreme cases, it might even prevent PBH production despite the presence of 
significant magnetic field. 
Taking into account the relevant physical effects on the magnetized overdense region, 
we establish an upper-limit on the amplitude of comoving magnetic fields,
approximately $0.13-0.15 {\rm \mu G}$ at a scale of $10^{17} {\rm Mpc}^{-1}$. 
Additionally, we compare our constraints with various small-scale probes.}

\begin{document}
\maketitle
\flushbottom

\section{Introduction}

Observational probes have confirmed the presence of magnetic fields in the universe at all length scales.
They have been observed in the galaxies and galaxy clusters with the typical strength of order micro-Gauss with a coherence length over kpc to Mpc~\cite{1994-Kronberg-Rept.Prog.Phys.,1996-Beck.etal-ARAA,2002-Widrow-Rev.Mod.Phys.,2004-Vallee-NAR,2004-Gaensler.Beck.etal-AstroRev,2004-Giovannini-IJMPD,2007-Barrow.etal-PhyRep,2011-Kandus.Kunze.Tsagas-PhysRept,2013-Durrer.Neronov-Arxiv,2016-Subramanian-Arxiv}. The origin of these magnetic fields is still unknown and they could have cosmological or astrophysical origin. The FERMI and High Energy Stereoscopic System (HESS) measurements of $\gamma$-rays from blazars provide a lower bound of the order of $10^{-16} \rm{G}$ in intergalactic voids~\cite{2010-Neronov.Vovk-Sci}. 
In the literature, various proposals argue that this lower bound suggests that these magnetic fields could be of primordial origin, which are generated in the early universe and later amplified due to dynamo mechanism~\cite{1988-Turner.Widrow-PRD,1991-Ratra-Apj.Lett,1991-Vachaspati-PLB,1993-Dolgov-PRD,2007-Martin.Yokoyama-JCAP,2016-Mukohyama-PRD,2019-Kushwaha.Shankaranarayanan-PRD,2020-Bamba.Odintsov.etal-JCAP,2021-Giovannini-JCAP,2021-Tripathy.etal-arXiv,2021-Nandi-JCAP,Papanikolaou:2023nkx,Papanikolaou:2023cku}. In most of these scenarios, either the magnetic field is too weak to be amplified via dynamo action or the coherence length is too small. In contrast to these proposals, in a recent study, it is shown that the primordial magnetic fields (PMFs) on large scales are ruled out from baryon isocurvature constraints~\cite{Kamada:2020bmb}.

The magnetic fields in the early universe affect the cosmological processes, leading to distinct signatures that constrain the properties of the PMFs.  For example, magnetic field energy density at the Big Bang Nucleosynthesis (BBN) epoch might increase ${\rm He}$ production beyond the observational bounds, which allows to constrain its comoving amplitude $ < 1.5 {\rm \mu G}$ independent of the coherence scale~\cite{2012-Kawasaki.Kusakabe-PRD}. Observations of the temperature and polarization anisotropies of CMB provide one of the most promising ways to detect or constrain the properties of PMFs. A homogeneous PMFs due to its preferred spatial direction would lead to anisotropic expansion around that direction, hence leading to a quadrupole anisotropy in the CMB, which puts an upper-limit on the amplitude of order nano-Gauss (on Mpc scales) at the present epoch~\cite{2013-Durrer.Neronov-Arxiv,2016-Subramanian-Arxiv}. Furthermore, before recombination, PMFs induce small-scale baryonic density fluctuations, leading to an inhomogeneous recombination process that affects the CMB anisotropies. Using magnetohydrodynamic (MHD) effects, authors in Ref.~\cite{2019-Jedamzik.Saveliev-PRL} put the most stringent upper-limit of $\lesssim 47 {\rm pG}$ (present day) for scale-invariant PMFs power spectra.

While most of the above-mentioned studies focus on explaining or constraining the properties of large-scale magnetic fields, small-scale PMFs have not received much attention. 
Given that the early universe was magnetized and the most favorable scenarios to explain the origin of the PMFs should occur before the recombination epoch, we can ask if there is a way to constrain the properties of PMFs on small scales without being affected by sub-Horizon scale processes such as MHD or dynamo action. 
By doing that, we can not only gain a better understanding of the nature of \emph{seed} magnetic field, but also constrain the early-time magnetogenesis scenarios.
Small-scale magnetic fields in the radiation-dominated (RD) era can lead to some interesting consequences; for example, they can create matter-antimatter asymmetry~\cite{2016-Kamada.Long-2,2016-Fujita.Kamada-PRD,2021-Kushwaha.Shankaranarayanan-PRD}, CMB distortion~\cite{2014-Kunze.Komatsu-JCAP}, magnetic reheating~\cite{2017-Saga.etal.Yokoyama-MNRAS},  gravitational wave background (GWB)~\cite{2018-Saga.etal-PRD} and might also affect the primordial black hole (PBH) formation~\cite{2020-Saga.Hiroyuki.Yokoyama-JCAP}. To constrain the upper-limit of the amplitude of small scale PMFs, we need to investigate the observable signatures of PMFs on small-scale cosmological processes, for example GWB and PBHs formation.

In this work, we present a systematic study to derive the upper-limit on the amplitude of PMFs by taking into account the physical effects affecting the formation of PBHs from magnetized overdense regions. This program can be achieved by using the fact that the magnetic fields surpassing
a critical threshold source the primordial curvature perturbation to a degree
that results in the over-production of PBHs, 
exceeding the observational upper limit on their abundance.
Upper limits on PMFs based on this consideration has been already given in the previous work~\cite{2020-Saga.Hiroyuki.Yokoyama-JCAP}. In the present paper, we point out a new effect which was
over-looked in the previous study: change of the threshold value of the PBH formation
by the pressure of the magnetic fields.
This effect acts as weakening the upper limits on the PMFs.
Yet, as it will turn out, our constraints are severer than those provided in Ref.~\cite{2020-Saga.Hiroyuki.Yokoyama-JCAP} because of another effect (i.e., temperature dependence of
the relativistic degrees of freedom) which is not included in the previous study.
To simplify the calculations and to make the direct comparison with the previous work, 
we consider the same power spectrum for the PMFs as the one
considered in \cite{2020-Saga.Hiroyuki.Yokoyama-JCAP}.
We also compare the constraints for both Gaussian and non-Gaussian type probability distribution functions of the density contrast. 

Before closing this section, we stress that our constraints are applicable as long as the PMFs were once super-Hubble scales in the RD era. Furthermore, our constraints will also provide a great insight into understanding the magnetogenesis scenarios and might rule out certain class of models. 

This work is organized in the following way: in section~\ref{sec:curvature_pert}, we discuss the evolution of curvature perturbations sourced by PMFs-induced anisotropic stress and calculate the density contrast. Section~\ref{sec:jeans-criterion} discusses the effects of magnetic fields on the overdense region, then in section~\ref{sec:PBH_formation}, we compute abundance of PBHs and \ref{sec:constraining_PMF} discusses our constraints on the PMFs and comparison with other observational probes. We work with the natural units where $\hbar = c = k_B = 1$ and the reduced Planck mass $M_{\rm Pl} = 2.43\times 10^{18} {\rm GeV}$. For the magnetic field, we follow the electromagnetic CGS units where $1 \rm{G} = 6.92\times 10^{-20} {\rm GeV^2}$.

\section{Anisotropic stress and density contrast sourced by PMFs}\label{sec:curvature_pert}

\subsection{Curvature perturbation from PMFs}
In this section, we discuss the evolution of primordial curvature perturbation sourced by the anisotropic stress of the PMFs.
We assume that the inhomogeneous magnetic fields are generated before the radiation-dominated era (for example, during inflation), with the energy density of the magnetic field being the first-order perturbation to the Friedmann–Lemaître–Robertson–Walker (FLRW) spacetime~\cite{2002-Mack.Kahniashvili.etal-PRD,2012-Suyama.Yokoyama-PRD}. 
Since the primordial plasma, which is produced by the decay of the inflaton after inflation,
is highly conducting, magnetic fields are frozen-in and decay as $1/a^2(\eta)$. This allows to separate the spatial and temporal dependence of the magnetic fields on sufficiently large-scales as 
$\textbf{B}(\textbf{x},\eta) = \textbf{B}(\textbf{x})/a^2(\eta)$, where $\eta$ is the conformal time, and we set the scale factor (in cosmic time $t$) at present epoch to be unity, $a(t_0) = 1$. The Fourier transform of the magnetic field is defined as:
\begin{equation}
\textbf{B}(\textbf{k}) = \int d^3x \textbf{B}(\textbf{x}) e^{-i \textbf{k}\cdot\textbf{x}} .
\end{equation}
Assuming $\textbf{B}(\textbf{k})$ is a Gaussian random field, the statistical properties of the field $\textbf{B}(\textbf{k})$ are determined by the two-point correlation function given by~\cite{2002-Mack.Kahniashvili.etal-PRD,2010-Shaw.Lewis-PRD,2013-Durrer.Neronov-Arxiv}:
\begin{align}\label{eq:two-point-correlation-Bk}
    \langle B_i(\textbf{k}) B_j(\textbf{k}^{\prime}) \rangle = (2\pi)^3 \delta^3 (\textbf{k} - \textbf{k}^{\prime}) \frac{P_{ij}(\hat{k})}{2} P_B (k) 
\end{align}
where $\hat{k}_i = k_i/k$ are unit wavenumber components, $\delta^3 (\textbf{k} - \textbf{k}^{\prime})$ is 3D Dirac-delta function. $P_{ij}(\hat{k}) = \delta_{ij} - \hat{k}_i \hat{k}_j$ is projection tensor that comes from the divergence condition $\nabla \cdot \textbf{B} = 0$ and $ P_B(k)$ is the magnetic power spectrum.

Now, let us consider the spatial part of the energy-momentum tensor due to the 
magnetic fields $T_{ij}$.
Following \cite{2010-Shaw.Lewis-PRD}, we express $T_{ij}$ as
$T_{ij} = p_\gamma ( g_{ij} \Delta_B + \Pi_{ij}^B )$, where $\Delta_B$ is the perturbations in energy density, $p_\gamma$ is the pressure of photons and
$\Pi_{ij}^B$, which sources the curvature perturbation through the mode-mode coupling, 
is the anisotropic stress due to magnetic fields which is traceless i.e., ${\Pi^B}^i_i = 0$.
The scalar part of the anisotropic stress in momentum space is given by~\cite{2010-Shaw.Lewis-PRD}:
\begin{align}\label{eq:Pi_k-def}
\Pi_B (\mathbf{k},T) & = -\frac{3}{2}  \left( \hat{k}^i \hat{k}^j - \frac{\delta^{i j}}{3} \right) \Pi^{ij}_B \nonumber \\
&=  \left( \hat{k}^i \hat{k}^j - \frac{\delta^{i j}}{3} \right) \frac{9}{8\pi \rho_{\gamma} (T) a^4(T)} \int \frac{d^3 \textbf{k}^{\prime} }{(2\pi )^3} B_i (\mathbf{k}^{\prime}) B_j (\mathbf{k} - \textbf{k}^{\prime})
\end{align}
and the spatial indices are raised and lowered by the Kronecker delta $\delta_{ij}$ and $\rho_{\gamma} (T)$ is the photon energy density which varies with temperature of the Universe.
Using the conservation of entropy i.e., $g_{*s} (T) a^3 (T) T^3 = \text{constant}$, 
we obtain
\begin{align}\label{eq:rho_gamma-T-relation}
     \rho_{\gamma} (T) = \rho_{\gamma,0} \left( \frac{a (T_0)}{a(T)} \right)^4  \left(\frac{g_{*s} (T_0) }{g_{*s} (T)} \right)^{4/3} 
\end{align}
where $\rho_{\gamma ,0}$ is the photon energy density at present epoch ($T_0$) and $g_{*s}$ is a number of relativistic degree of freedom. Thus, using Eq.~\eqref{eq:rho_gamma-T-relation} in Eq.~\eqref{eq:Pi_k-def} gives the scalar part of the anisotropic stress as
\begin{align}\label{eq:Pi_B-final}
    \Pi_B (\textbf{k},T) = \left( \hat{k}^i \hat{k}^j - \frac{\delta^{i j}}{3} \right) \frac{9}{8\pi \rho_{\gamma,0}}  \left(\frac{g_{*s} (T) }{g_{*s} (T_0)} \right)^{4/3}  \int \frac{d^3 \textbf{k}^{\prime} }{(2\pi )^3} B_i (\mathbf{k}^{\prime}) B_j (\mathbf{k} - \textbf{k}^{\prime})  .
\end{align}
Note that the above expression differs from  Ref.~\cite{2020-Saga.Hiroyuki.Yokoyama-JCAP} wherein the dependence on relativistic degrees of freedom is missing.
The two-point correlation function of the scalar part of the anisotropic stress can be obtained as
\begin{align}\label{eq:piB-piB-final}
    \langle \Pi_B (\textbf{k},T) \Pi_B^* (\textbf{p},T) \rangle &= \delta^3 (\textbf{k}-\textbf{p})  \left( \frac{9}{16\pi \rho_{\gamma,0}^2} \right)  \left(\frac{g_{*s} (T) }{g_{*s} (T_0)} \right)^{8/3}
    \int d k^{\prime} {k^{\prime}}^2 P_B (k^{\prime}) \int_{-1}^{1} d\gamma P_B (|\textbf{k} -\mathbf{k}^{\prime}|)
     \nonumber\\
     &{} \qquad \qquad \times \left[ \frac{k^2 (1+\gamma^2) + k k^{\prime} (2 \gamma - 6\gamma^3 ) +  {k^{\prime}}^2 (5 + 9\gamma^4 - 12\gamma^2) }{4(k^2+ {k^{\prime}}^2-2\gamma k k^{\prime})} \right]
\end{align}
where $\gamma = \hat{k}\cdot\hat{k}^{\prime}$. 

The anisotropic stress due to primordial magnetic fields acts as a source for the curvature perturbation. In particular, it induces two kinds of contributions to the primordial curvature perturbations \cite{2010-Shaw.Lewis-PRD}: (i) \emph{passive modes}, which are induced before neutrino decoupling on both super-horizon scales and sub-horizon scales, and (ii) \emph{compensated modes}, which are induced only on sub-horizon scales. In this work, we will only focus on passive modes because only super-horizon modes are relevant for the formation of PBHs in the radiation-dominated era. Note that after neutrino decoupling time $\eta_{\nu}$, the anisotropic stress due to free streaming neutrinos compensates for primordial magnetic fields. Therefore, the significant growth of the PMFs sourced curvature perturbation $\zeta_B (\textbf{k},\eta)$ occur between the time of magnetic field generation $\eta_B$ and neutrino decoupling time $\eta_{\nu}$ on super-horizon scales. During this growing regime, the PMF-induced primordial curvature perturbation on super-horizon scales ($k\eta \ll 1$) on comoving slices is given by~\cite{2010-Shaw.Lewis-PRD,2013-Durrer.Neronov-Arxiv,2016-Subramanian-Arxiv}
\begin{align}\label{eq:curvature_pert_B}
    \zeta_B (\textbf{k},\eta) = -\frac{1}{3} \xi (\eta ) R_{\gamma} (\eta) \Pi_B (\textbf{k},\eta) 
\end{align}
where 
\begin{eqnarray}\label{eq:xi-expression}
\xi(\eta) =
\begin{cases}
 \log \left( \frac{\eta}{\eta_B} \right) + \frac{\eta_B}{2 \eta} - \frac{1}{2} & \qquad (\eta_B < \eta < \eta_\nu) \\
 \log \left( \frac{\eta_\nu}{\eta_B} \right) + \left( \frac{5}{8 R_\nu} - 1 \right) & \qquad (\eta > \eta_\nu) 
\end{cases}
\end{eqnarray}
and $R_{\gamma} (\eta) = \frac{\rho_{\gamma}}{\rho_{\rm tot}}$ refers to the energy fraction of the photon component in the total radiation. 
As we can see the passive mode depends logarithmically on the magnetic field generation epoch $\eta_B$ which in our case is determined by the scale of inflation. The dimensionless power-spectrum $\mathcal{P}_{\zeta_B} (k,\eta)$ is related to the two-point correlation function of comoving curvature perturbation as
\begin{align}\label{eq:zeta-zeta-def}
    \langle \zeta_B (\textbf{k},\eta) \zeta_B^* (\textbf{p},\eta)  \rangle = (2\pi)^3 \delta^3 (\textbf{k} - \textbf{p}) \frac{2\pi^2}{k^3}\mathcal{P}_{\zeta_B} (k,\eta) .
\end{align}
Using Eq.\eqref{eq:piB-piB-final} and Eq.\eqref{eq:curvature_pert_B}, we can calculate the two-point correlation function of the comoving curvature perturbation as
\begin{align}\label{eq:zeta-zeta-expression}
    \langle \zeta_B (\textbf{k},\eta) \zeta_B^* (\textbf{p},\eta)  \rangle &=  \frac{1}{9} \xi^2 (\eta ) R_{\gamma}^2 (\eta ) \langle  \Pi_B (\textbf{k},\eta) \Pi_B^* (\textbf{p},\eta)  \rangle
    \nonumber\\
    &= \delta^3 (\textbf{k}-\textbf{p})  \frac{1}{9} \xi^2 (\eta ) R_{\gamma}^2 (\eta )  \left( \frac{9}{16\pi \rho_{\gamma,0}^2} \right)  \left(\frac{g_{*s} (T) }{g_{*s} (T_0)} \right)^{8/3}
    \int d k^{\prime} {k^{\prime}}^2 P_B (k^{\prime}) \int_{-1}^{1} d\gamma P_B (|\textbf{k} -\mathbf{k}^{\prime}|)
     \nonumber\\
     &{} \qquad \qquad \times \left[ \frac{k^2 (1+\gamma^2) + k k^{\prime} (2 \gamma - 6\gamma^3 ) +  {k^{\prime}}^2 (5 + 9\gamma^4 - 12\gamma^2) }{4(k^2+ {k^{\prime}}^2-2\gamma k k^{\prime})} \right]
\end{align}
By comparing Eq.\eqref{eq:zeta-zeta-def} and Eq.\eqref{eq:zeta-zeta-expression}, we obtain the dimensionless power-spectrum for the passive comoving curvature perturbation mode as
\begin{align}\label{eq:power-spectrum-zetaB}
    \frac{2\pi^2}{k^3}\mathcal{P}_{\zeta_B} (k,\eta) &=  
     \frac{1}{(2\pi)^3} \xi^2 (\eta ) R_{\gamma}^2  \left( \frac{1}{16\pi \rho_{\gamma,0}^2} \right)  \left(\frac{g_{*s} (T) }{g_{*s} (T_0)} \right)^{8/3}
    \int d k^{\prime} {k^{\prime}}^2 P_B (k^{\prime}) \int_{-1}^{1} d\gamma P_B (|\textbf{k} -\mathbf{k}^{\prime}|)
     \nonumber\\
     &{} \qquad \qquad \times \left[ \frac{k^2 (1+\gamma^2) + k k^{\prime} (2 \gamma - 6\gamma^3 ) +  {k^{\prime}}^2 (5 + 9\gamma^4 - 12\gamma^2) }{4(k^2+ {k^{\prime}}^2-2\gamma k k^{\prime})} \right]  .
\end{align}
Apart from the factor accounting for the change of the relativistic degrees of freedom,
this result recovers the one given in \cite{2020-Saga.Hiroyuki.Yokoyama-JCAP}.
Eq.~(\ref{eq:power-spectrum-zetaB}) will be used for estimating the variance of the magnetized overdense region, which determines the properties of the PBH formation.

\subsection{Density contrast and its variance}
The curvature perturbation $\zeta_B$ is not a quantity that directly represents the 
local (=horizon-scale) spatial curvature. 
For the purpose of computing the PBH abundance, the density contrast on the comoving slice 
$\delta_B$ is more suitable quantity than $\zeta_B$ \cite{Young:2014ana}.

In the radiation-dominated epoch, the relation between comoving curvature perturbation $\zeta_B$ and density contrast $\delta_B (\eta,k)$ is given by~\cite{2004-Green.etal-PRD}:
\begin{align}\label{eq:delta_B-zeta_B}
    \delta_B (\textbf{k},\eta) = \frac{4}{9} \left( \frac{k}{a H} \right)^2 \zeta_B (\textbf{k},\eta) =
     - \frac{4}{27} \left( \frac{k}{a H} \right)^2  \left[ \log \left(\frac{\eta}{\eta_B}\right) + \frac{\eta_B}{2\eta} - \frac{1}{2} \right] R_{\gamma}(\eta) \Pi_B (\textbf{k},\eta)
\end{align}
where we have used Eq.~\eqref{eq:curvature_pert_B} in obtaining the last equality.
The smoothed density perturbation over the comoving radius $R$ is given by
\begin{align}
    \delta_B (\textbf{x},\eta;R) = \int d^3 x^{\prime} W_R (|\textbf{x} - \textbf{x}^{\prime}|) \delta_B (\textbf{x}^{\prime},\eta) = \int \frac{d^3 k}{(2\pi)^3} W (kR) \delta_B (\textbf{k},\eta) e^{i\textbf{k}\cdot\textbf{x}}
\end{align}
where $W(kR) = e^{-k^2 R^2/2} $ is the Window function.
In calculating the PBH abundance, the smoothing scale $R$ is usually taken to be the comoving horizon scale at the PBH formation~\cite{2018-Sasaki.etal-CQG,2020-Carr.Florian-arXiv}. 
The variance is related to the power spectrum of the density perturbation by the relation~\cite{2016-Nakama.Suyama-PRD}:
\begin{align}\label{eq:sigma_B-2}
    \sigma^2_{B} (R,\eta) \equiv \langle \delta_B (\textbf{x},\eta ; R)^2 \rangle - \langle \delta_B (\textbf{x},\eta ; R) \rangle^2 = \int \frac{dk}{k} W^2 (kR) \mathcal{P}_{\delta_B} (k,\eta),
\end{align}
where $\mathcal{P}_{\delta_B} (k,\eta )$ is the dimensionless
power spectrum of $\delta_B$.
The fact that magnetic fields are divergenceless, and from the  Eqns.(\ref{eq:two-point-correlation-Bk}, \ref{eq:Pi_B-final}), the mean value of the curvature perturbation induced by PMFs vanishes in the real space $\langle \zeta_B (\textbf{x})\rangle = 0$. Thus, from Eq.\eqref{eq:sigma_B-2}, we see that the mean density perturbation is zero, $  \langle \delta_B (\textbf{x},\eta ; R) \rangle=0$.
Using \eqref{eq:delta_B-zeta_B}, the variance of $\delta_B$ can be obtained as
\begin{align}\label{eq:sigma_B-final}
    \sigma^2_{B} (R,\eta)  &= \frac{16}{81} \int \frac{dk}{k} W^2 (kR)  (k\eta)^4 \mathcal{P}_{\zeta_B} (k,\eta)
\end{align}
We would like to mention that up to this point, the analysis is generic in the sense that we have not used any specific form for the magnetic power spectrum. However, to study the effect of the magnetic fields on the formation of PBHs, we need to consider the magnetic field power spectrum. We will discuss this in section~\ref{sec:PBH_formation}.

\section{Effects of PMF on overdense region}\label{sec:jeans-criterion}

In this section, we discuss the effect of PMFs on the overdense region and study the consequences which are particularly important for PBH formation.

\subsection{Effect of the magnetic fields
on the Jeans length}
To determine the PBH abundance from the given variance of the density contrast (i.e., 
Eq.~\eqref{eq:sigma_B-final}),
knowledge of the threshold of the PBH formation is needed.
In the case of the standard scenario where PBH forms purely from perturbation of
radiation, the threshold has been investigated in detail both analytically and numerically~\cite{1975-Carr-ApJ, Shibata:1999zs, 2005-Musco.Miller.Rezzolla-CQG, 2013-Harada.etal-PRD, Nakama:2013ica, Musco:2018rwt, Young:2019osy, Escriva:2019nsa}.
The simplest version, which was derived in \cite{1975-Carr-ApJ} based on the Newtonian gravity, 
gives the threshold value $\delta_{\rm th}=c_s^2 (=\frac{1}{3})$ 
and it still serves as a physically reasonable estimate of threshold at least qualitatively.
In what follows, we provide a physical argument that presence of the magnetic field 
changes the threshold and derive the correction to the threshold 
coming from the magnetic field based on the Jeans criterion in Newtonian dynamics.
General relativistic effect will change our result, but we believe that our correction
captures the essential features of the effect caused by the magnetic field and 
is valid within a factor of ${\cal O}(1)$ uncertainty.

According to Jeans criterion, the perturbations (amplitude) would grow (against the radiation pressure) if the size of the perturbed region is greater than the Jeans length $\lambda_J$. 
If the uniform magnetic field pervades the medium (which is assumed to be highly conducting), the dispersion relation for the perturbations changes depending on the angle between the direction of propagation of perturbations and the magnetic field vector $\textbf{B}$. For example, the dispersion relation for the modes propagating perpendicular to the direction the magnetic field is given by~\cite{1961-Chandrasekhar-Book} 
\begin{align}\label{eq:modifiedDR-Bfield}
    \omega^2 = \frac{B^2 k^2}{4 \pi \rho} + c_s^2 k^2 - 4 \pi G \rho
\end{align}
where $B$ is the magnetic field strength, $c_s$ is the sound speed, $\rho$ is the energy density of the Universe. 
Gravitational instability occurs for modes having negative $\omega^2$.
Using the modified dispersion relation in Eq.\eqref{eq:modifiedDR-Bfield}, we can obtain the modified Jeans length referred to as \emph{magnetic Jeans length} $\lambda_B$, which for the modes traveling perpendicular to the direction of the magnetic field is given by~\cite{1961-Chandrasekhar-Book,1966-Strittmatter-MNRAS}
\begin{align}\label{eq:magnetic_jean_length}
    \lambda_B \equiv 
     \lambda_J \, \sqrt{ 1 + \frac{2 \rho_B}{ c_s^2 \rho }  }  = \lambda_J \, \sqrt{ 1 + \frac{2 r_B}{ c_s^2}  } ,
\end{align}
where $r_B = \rho_B /\rho$ measures the fraction of magnetic field energy in a given patch. In what follows, we adopt this equation as the representative 
Jeans length in the presence of the magnetic field.
This is in sharp contrast to Ref.~\cite{2020-Saga.Hiroyuki.Yokoyama-JCAP} in which
the effect of the magnetic field on the Jeans length was neglected.  
To the best of our knowledge, this paper is the first to
incorporate the effect of the magnetic field as $\lambda_B$ defined above
in the context of PBH formation.
Note that the additional term in the above equation is similar to the effect of rotation on the Jeans length discussed in~\cite{2019-He.Suyama-PRD}.
As we can see, the magnetic field increases the Jeans length.
For instance, for $r_B=0.1$, we have $\left. \lambda_B\right|_{r_B = 0.1} = 1.265 \lambda_J$.
Although the increase is only by $26\%$, this does not necessarily mean
that the impact of the magnetic field on the PBH abundance is mild as well
since the abundance is typically exponentially sensitive to the Jeans length.
We will later discuss this effect in detail in section~\ref{sec:PBH_formation}. 

\subsection{Estimation of threshold density contrast in magnetized environment}

In the previous subsection, we have shown that the presence of a uniform magnetic field affects the Jeans criterion by increasing the Jeans length~\eqref{eq:magnetic_jean_length} by a factor that depends on the magnetic field strength. 
To estimate the effect of the magnetic field on the threshold density contrast, we first briefly review
how the threshold is derived in the standard case (i.e., no magnetic field) \cite{1975-Carr-ApJ}.
To this end, let us consider the overdense region with the evolution given by Friedmann equation for the closed Universe~\cite{2018-Sasaki.etal-CQG,2020-Carr.Florian-arXiv}
\begin{align}\label{eq:FriedmannEq1}
    H^2 = \frac{8\pi G}{3} \rho - \frac{1}{a^2}
\end{align}
where $\rho \equiv \bar{\rho} (1+\delta)$ with $\bar{\rho}$ being the unperturbed background energy density of the Universe and $\delta \equiv (\rho - \bar{\rho})/\rho = \delta \rho/\rho$ is the density contrast. 
Because of the curvature term, the overdense region turns from the expansion phase to the contraction phase; we refer to this turnaround as the maximum expansion and add the subscript ``max" to any quantities evaluated at that time. The time when the proper size of the overdense region becomes equal to the Hubble horizon (horizon re-entry) is denoted by ``hc" which refers to horizon-crossing. The proper size of the overdense region at this maximum expansion time is $\left( a_{\rm max}/a_{\rm hc} \right) H_{\rm hc}^{-1}$ and the Jeans length evaluated at the maximum expansion time is given by:
\begin{align}\label{eq:jeans_length}
    \lambda_J = \frac{c_s}{\sqrt{G\rho_{\rm max}}} = \frac{c_s}{\sqrt{G\bar{\rho}_{\rm max} (1+\delta_{\rm max}) }} \simeq c_s a_{\rm max} .
\end{align}
From Eq.\eqref{eq:FriedmannEq1}, we have 
$a_{\rm max} = \sqrt{\frac{3}{8\pi G \rho_{\rm max}}}$. 
Imposing the Jeans criterion at maximum expansion time gives the condition for the formation of PBHs
\begin{align}\label{eq:jeans_criterion}
    \frac{a_{\rm max}}{a_{\rm hc}} \frac{1}{H_{\rm hc}} > \lambda_J \simeq c_s a_{\rm max}.
\end{align}
The relation between $a_{\rm max}$ and $a_{\rm hc}$ in terms of density contrast at horizon crossing is given by
\begin{align}\label{eq:a_max-a_hc-delta_hc-relation}
    \frac{a_{\rm max}}{a_{\rm hc}} = \sqrt{ \frac{1+\delta_{\rm hc}}{\delta_{\rm hc}}  },
\end{align}
and constancy of $\rho (1+\delta) a^4$ gives
\begin{align}\label{eq:a_max-H_hc-delta_hc-relation}
    H_{\rm hc} a_{\rm max} = \frac{\sqrt{1+\delta_{\rm hc}}}{\delta_{\rm hc}}.
\end{align}
Plugging these relations in Jeans criterion~\eqref{eq:jeans_criterion}, we obtain the threshold density contrast~\cite{1975-Carr-ApJ}
\begin{align}\label{eq:jean_criterion-delta_hc}
    \delta_{\rm hc} > \delta_{\rm th} = c_s^2 .
\end{align}

Now, let us include the presence of the magnetic fields in the above picture, 
and investigate how the presence of uniform magnetic fields in the Universe changes the threshold density contrast~\eqref{eq:jean_criterion-delta_hc}. As we have discussed in the previous subsection, the presence of the uniform magnetic fields modifies the Jeans length which is given by Eq.\eqref{eq:magnetic_jean_length}. Note that even in the presence of uniform magnetic fields, the original argument that the size of the overdense region at the maximum expansion time needs to be greater than the Jeans length to form a black hole should still hold. 
This means the condition for the PBH formation in the presence of the 
magnetic field is obtained by replacing the original Jeans length with $\lambda_B$. 
Following the calculations as done previously but with $\lambda_B$ on RHS of Eq.\eqref{eq:jeans_criterion}, we obtain
\begin{align}\label{eq:modified_jeans_criterion-1}
     \frac{a_{\rm max}}{a_{\rm hc}} > c_s (a_{\rm max} H_{\rm hc} ) \sqrt{ 1 + \frac{2 \rho_B}{ c_s^2 \rho }  }
\end{align}
where we used $\lambda_J = c_s a_{\rm max}$. Now, using Eq.\eqref{eq:a_max-a_hc-delta_hc-relation}
 and Eq.\eqref{eq:a_max-H_hc-delta_hc-relation} in Eq.\eqref{eq:modified_jeans_criterion-1}, we can calculate the threshold density contrast for the formation of PBHs in the presence of uniform magnetic field as
\begin{align}\label{eq:magnetic_density_contrast}
    \delta^B_{\rm hc} > c_s^2 \left( 1 + \frac{2r_B}{ c_s^2 }   \right)  \quad
    \implies \quad \delta_{\rm th}^B = c_s^2 + 2 r_B. 
\end{align}
From Eq.~\eqref{eq:magnetic_density_contrast}, we can see that a sufficiently strong magnetic field significantly enhances the threshold density contrast than the non-magnetized case. This can be understood as follows: a sufficiently strong magnetic field would increase the total pressure due to an extra contribution from magnetic pressure, which would act against the gravitational attraction. Therefore, higher density contrast is required to form PBHs from the gravitational collapse of the magnetized overdense region.

Before closing this section,
we would like to mention that the threshold (\ref{eq:magnetic_density_contrast}) 
has been derived analytically based on the simple Newtonian picture and 
it is merely approximate. 
According to general relativistic simulations (without the magnetic field) ~\cite{2005-Musco.Miller.Rezzolla-CQG}, 
the threshold is not only different from $c_s^2$ but also varies in 
the range $0.4 - 0.5$ depending on the density profile of the overdense region.
We expect the contribution of the magnetic field, which is given as $2r_B$ in (\ref{eq:magnetic_density_contrast}), also possesses such ${\cal O}(1)$
variations in the realistic cases.
Yet, since there are no simulations of PBH formation in the presence of the
magnetic field, to keep consistency of our analysis, we use 
the threshold (\ref{eq:magnetic_density_contrast}) in the following analysis.

\section{Computation of PBH abundance}\label{sec:PBH_formation}

In this section, we provide a formalism to compute the PBH abundance in the context
of our scenario.

\subsection{Spiky power spectrum of PMF}
In what follows, for simplicity, we consider a Delta-function type power spectrum given by
\begin{align}\label{eq:ps-delta_fn}
    P_B (k) = \frac{2\pi^2}{k^3} B_p^2 k_p \delta (k-k_p)
\end{align}
where amplification is peaked at $k=k_p$ and $B_p$ is the comoving magnetic field at the scale $k=k_p$. 
Physically, $B_p$ represents the amplitude of the virtual magnetic field 
at present time obtained by simply evolving the primordial magnetic field in
an adiabatic manner throughout the whole redshift range.
In reality, magneto-hydrodynamics in general completely changes the 
evolution of the magnetic field once the scale $k_p$ enters the eddy turnover scale,
and $B_p$ does not represent the actual magnitude of the present-day magnetic field.
Nevertheless, in order to make a direct comparison with the previous work \cite{2020-Saga.Hiroyuki.Yokoyama-JCAP}, we use this quantity to place the upper limit
on the {\it primordial} magnetic field.

Magnetic power spectrum with a sharp spike is realized in the models where magnetic fields are generated due to the parametric resonance mechanism~\cite{2012-Byrnes.etal-JCAP,2020-Patel.etal-JCAP,2022-Sasaki.Vardanyan.etal-arXiv}. In these mechanisms, the large amplification of EM modes happens only for the specific modes lying in the resonance band determined by peak wavenumber $k_p$ and the model's parameter space.
For the Delta-function type magnetic field power-spectrum~\eqref{eq:ps-delta_fn}, the dimensionless power-spectrum for the passive comoving curvature perturbation~\eqref{eq:power-spectrum-zetaB} becomes
\begin{align}\label{eq:power-spectrum-zetaB-3}
     \mathcal{P}_{\zeta_B} (k,\eta) &= \frac{1}{(2\pi)^3} \xi^2 (\eta ) R_{\gamma}^2  \left( \frac{1}{32\pi \rho_{\gamma,0}^2} \right)  \left(\frac{g_{*s} (T) }{g_{*s} (T_0)} \right)^{8/3}
     \\
     \nonumber 
     & \qquad \times \frac{\pi^2 k^2  B_p^4  \left( k^4 - 16 k^2 k_p^2 + 80 k_p^4\right) \theta (2k_p - k)}{16 k_p^6}  
\end{align}
Using the above Eq.\eqref{eq:power-spectrum-zetaB-3} in Eq.~\eqref{eq:sigma_B-final}, we can obtain the variance as
\begin{align}\label{eq:sigma_B}
   \sigma_B^2 (R,\eta) &=  \frac{B_p^4  }{5184 \pi^2  \rho_{\gamma,0}^2} \left(\frac{g_{*s} (T) }{g_{*s} (T_0)} \right)^{8/3} \xi(\eta)^2 R_{\gamma}^2 \left(\frac{\eta}{R} \right)^4 \left(\frac{e^{-4 (k_p R)^2}}{(k_p R)^6} \right) 
   \nonumber \\
   & \times\left( e^{4 k_p^2 R^2} (3 - 12 k_p^2 R^2 + 20 k_p^4 R^4) + 4 (k_p R)^4 - 16 (k_p R)^6 - 64 (k_p R)^8 - 3 
   \right).
\end{align}
At the time of the horizon reentry at which $k_p=aH$,
the variance of $\delta_B$ smoothed at the scale $R=\eta$ becomes
\begin{align}\label{eq:sigmaB_xiR-simplified}
   \sigma_B^2 (R,\eta=R) 
   \simeq 1.1 \times 10^{-3} {\left( \frac{B_p}{1~{\rm \mu G}} \right)}^4 
   \left(\frac{g_{*s} (\eta) }{g_{*s} (\eta_0)} \right)^{8/3}  (\xi(\eta) R_{\gamma}(\eta) )^2. 
\end{align}
From the above equation, we see that due to the function $\xi (\eta) =  \log \left(\frac{\eta}{\eta_B}\right) + \frac{\eta_B}{2\eta} - \frac{1}{2}$, the variance $\sigma_B^2 (R,\eta)$ depends on the epoch of magnetic field generation $\eta_B$.
In what follows, we take $T_B = 10^{14} {\rm GeV}$. 
We would like to emphasize that the variance $\sigma_B^2 (R,\eta)$ also depends significantly on the variation of relativistic degrees of freedom. To our knowledge, this effect is not discussed in the literature in the estimation of $\sigma_B^2 (R,\eta)$.
\begin{figure}[t]
\centering
\includegraphics[height=1.8in]{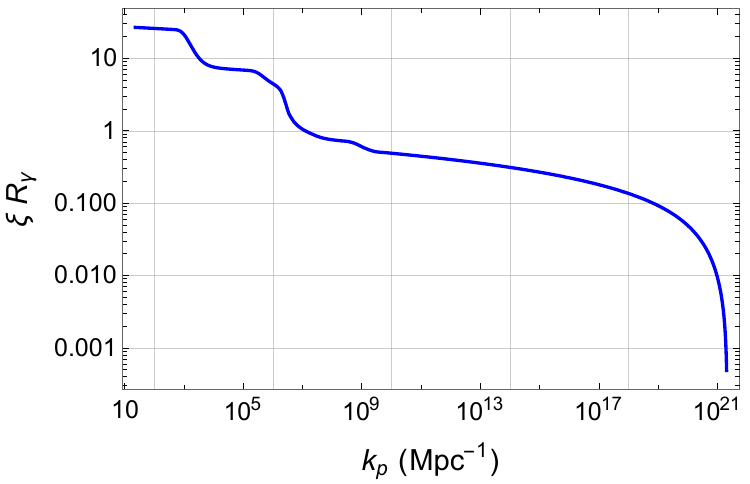} 
\includegraphics[height=1.8in]{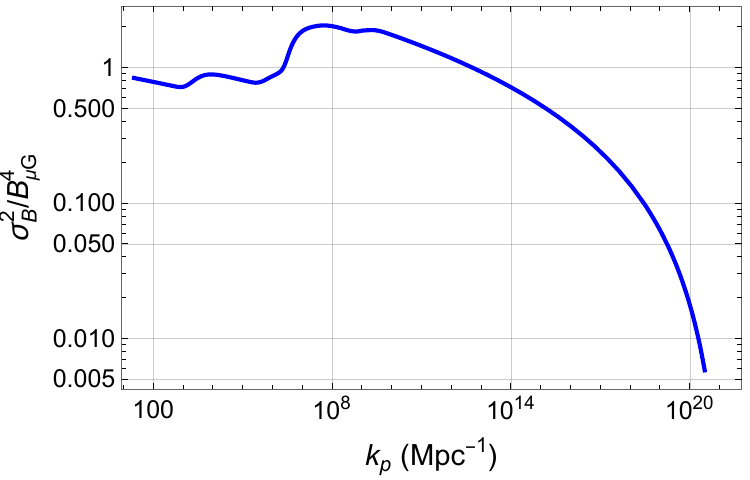} 
\caption{The behaviour of $\xi R_{\gamma}$ (left) and the variance $\sigma_B^2/B_{\rm \mu G}^4$ (right) with respect to $k_p$.}
\label{fig:sigma_B_plots}
\end{figure}

The left panel of \ref{fig:sigma_B_plots} shows $\xi (\eta) R_{\gamma}(\eta)$
as a function of $k_p$, where the evaluation time is taken to be the horizon crossing ($k=aH$). 
The right panel shows the variance of the density contrast smoothed over the horizon scale $R\sim (aH)^{-1}=\eta$ evaluated at the horizon crossing. 
This behaviour is not seen in Ref.~\cite{2020-Saga.Hiroyuki.Yokoyama-JCAP} as they consider the constant value of the quantity $\xi R_{\gamma}$. However, in this work, we incorporated the variation of relativistic degrees of freedom with temperature by using the data provided in Ref.~\cite{2018-Saikawa.Shirai-JCAP}
\subsection{PBH abundance}
To investigate the abundance of PBHs,
we define the parameter $\beta$, representing the mass fraction (energy density fraction) of PBHs at the time of their formation~\cite{2018-Inomata.etal-PRD,2018-Sasaki.etal-CQG}:
\begin{align}
    \left. \beta = \frac{\rho_{\rm PBH}}{\rho_{\rm tot}} \right|_{\rm form}. 
\end{align}
Denoting $P(\delta)$ by the probability distribution function of $\delta_B (R,R)$,
the fraction $\beta$ can be written as
\begin{equation}
\label{beta-general}
\beta =\gamma \int_{\delta> \delta^B_{\rm th}} P(\delta) d\delta,
\end{equation}
where $\gamma \sim 0.2$ is the numerical factor which appears due to the effect
that the PBH mass is not equal to the horizon mass at the time of the
horizon re-entry \footnote{ In the literature, often the upper limit of integration in Eq.~\eqref{eq:beta-relation-B-magJeans} is taken as $\delta_{\rm max}=\infty$ (see Ref.\cite{2020-Carr.Florian-arXiv,2023-LISACosmologyWorkingGroup}). 
However, both give approximately the same result but for $\delta_{\rm max} = 1$ case, we need to restrict $r_B$ to avoid $\delta_{\rm th}^B$ becoming greater than unity which leads to the negative value of $\beta$ (unphysical).}.
The lower limit of integration is non-trivial because $\delta_{\rm th}^B$ itself depends
on $\delta_B$, which means the lower limit is obtained as a solution of the algebraic equation 
\begin{equation}\label{eq:deltaB-cs-rB}
\delta_B =c_s^2+2 r_B(\delta_B).
\end{equation}
It is hard to precisely evaluate the second term on the right-hand side.
Rather than evaluating it precisely, which may require simulations and is beyond the
scope of our study,
we make an assumption 
\begin{equation}
    r_B \sim \frac{\delta_B}{\sigma_B} \bar{r}_B.
\end{equation}
Here ${\bar r_B}$ is the average value of $r_B$.
Given that both $\delta_B$ and $r_B$ depend quadratically on the magnetic field strength,
we expect this simple scaling provides approximate value of $r_B$ for a given $\delta_B$.
Substituting this ansatz to Eq.~(\ref{eq:deltaB-cs-rB}), the equation
becomes a simple linear equation for $\delta_B$. 
Solving it for $\delta_B$, we obtain
\begin{equation}
\label{eq:threshold}
\delta_B=\frac{c_s^2}{\left( 1-2\frac{\bar r_B}{\sigma_B} \right)}.
\end{equation}
To evaluate the right-hand side, we need to know $\sigma_B$ and ${\bar r_B}$.
We use Eq.~(\ref{eq:sigmaB_xiR-simplified}) to evaluate $\sigma_B$,
and ${\bar r_B}$ is given in the appendix \ref{app-rB-calc}.
Using these results, the above threshold becomes
\begin{equation}
\label{threshold-mag}
\delta_B =\frac{c_s^2}{\left( 1-\frac{5.8}{\xi (\eta)} \right)}.
\end{equation}
We adopt the right-hand side of this equation as the lower limit of the integration (\ref{beta-general}).

In Ref.~\cite{2021-Carr.etal-ReptProgPhys}, constraints on the abundance of PBHs over 
different mass ranges are provided. 
Applying those constraints to the fraction (\ref{beta-general}) in our scenario,
we can place upper limit on the amplitude of the
primordial magnetic fields $B_p$ on the scale $k_p$ by making a correspondence 
between $k_p$ and the PBH mass $M$ given as
\begin{align}\label{eq:Mpbh-k}
   k_p \simeq 7.5 \times 10^5   \gamma^{1/2} \left( \frac{g_*}{10.75} \right)^{-1/6}  \left( \frac{M}{30 M_{\odot} } \right)^{-1/2}  \rm{Mpc}^{-1} .
\end{align}
For the Gaussian distribution function of $\delta_B (R,R)$, the fraction
is given by~\cite{2018-Sasaki.etal-CQG}: 
\begin{align}\label{eq:beta-relation-B-magJeans}
    \beta (M) &= \gamma \int_{\delta^B_{\rm th}}^{1} d\delta ~~ \frac{1}{\sqrt{2\pi} \sigma_B} e^{-\frac{\delta^2}{2\sigma^2_B}}
    \simeq \frac{\gamma}{\sqrt{2\pi}} \left( \frac{\sigma_B}{\delta_{\rm th}^B} \right) \text{exp} \left(-\frac{ {\delta_{\rm th}^B}^2}{2\sigma_B^2}\right).
\end{align}
In reality, however, the density contrast is highly non-Gaussian since it 
depends quadratically on the magnetic field which is Gaussian.
Therefore, to 
correctly estimate PBH abundance, 
it is important to consider the non-Gaussian probability distribution function (PDF). In this work, we use the non-Gaussian distribution function studied in Ref.~\cite{2020-Saga.Hiroyuki.Yokoyama-JCAP}, which is given by:
\begin{align}\label{eq:rescaled_NG_PDF}
    P_{\rm NG} (x)= C \exp \left[ -1.5 \sqrt{1+ \frac{x^2}{b^2}} ~~\right]  
\end{align}
where $C$ and $b$ are free parameters which can be fixed by using the following conditions
\begin{align}\label{eq:conditions}
\int_{-\infty}^{\infty} P_{\rm NG} (x) dx= 1
\qquad
\int_{-\infty}^{\infty} x^2 P_{\rm NG} (x) dx= \sigma_{B}^2,
\end{align}
where $\sigma_{B}^2$ is given by Eq.~\eqref{eq:sigmaB_xiR-simplified}, 
which incorporates the effects of variation of relativistic degrees of freedom.
Using the above conditions gives
\begin{align}\label{eq-C-and-b}
C = 2.13 \sigma_B^{-1} ~, \qquad b =  0.84 \sigma_B.
\end{align}
In the next section, we show and compare the constraints on the amplitude of the PMFs for the non-Gaussian distribution function with the Gaussian distribution and also compare our constraints with Ref.~\cite{2020-Saga.Hiroyuki.Yokoyama-JCAP}.

\section{Constraints on the strength of PMFs from the abundance of PBHs}\label{sec:constraining_PMF}
\begin{figure}[t]
\centering
\includegraphics[height=3.4in]{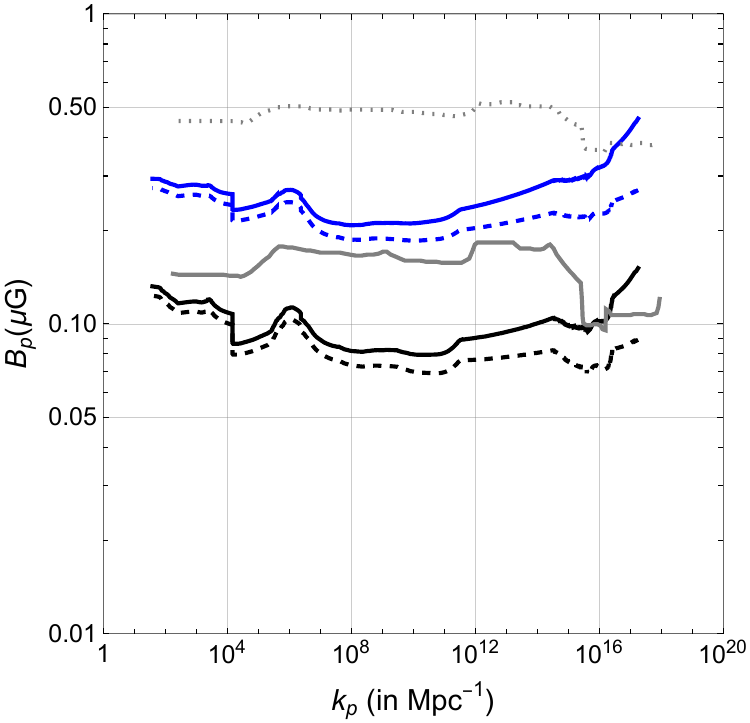} 
\caption{The constraints on $B_p$ for Gaussian (blue) and non-Gaussian (black) PDFs, dashed curves are for $\delta_{\rm th}$ and solid curves are for $\delta_{\rm th}^B$ which includes the effect of magnetic fields in Jeans length. Constraints from Ref.~\cite{2020-Saga.Hiroyuki.Yokoyama-JCAP} are shown in gray curves, where solid curve is for non-Gaussian and dotted curve is for Gaussian PDFs. }
\label{fig:B0_constraints_Gaussian-NG-compare}
\end{figure}
Having discussed the effects of magnetic fields on the PBH abundance for both types of distribution functions, our aim is now to constrain the amplitude of PMFs in the RD era. To do that, we present the constraints in terms of the comoving magnetic field amplitude $B_p$. 

As we mentioned in the previous section,
we employ the result provided in Ref.~\cite{2021-Carr.etal-ReptProgPhys} 
as the observational upper limits on the PBH fraction $\beta$ \footnote{
In Ref.~\cite{2021-Carr.etal-ReptProgPhys}, the upper limits are 
given in terms of
the rescaled fraction $\beta'$ defined by
\begin{align}\label{eq:beta_prime_def}
\beta^{\prime} (M) \equiv \gamma^{1/2} \left( \frac{g_*}{106.75} \right)^{-1/4} \left( \frac{h}{0.67} \right)^{-2} \beta (M) .
\end{align}
}.
\ref{fig:B0_constraints_Gaussian-NG-compare} shows the constraints on the amplitude of PMFs for both the Gaussian probability distribution function
and the non-Gaussian one given by Eq.~\eqref{eq:rescaled_NG_PDF}.
We find that inclusion of the effect of magnetic fields in the estimation of threshold density contrast ($\delta_{\rm th}^B$) affects the constraints for both Gaussian and non-Gaussian PDFs
in the same manner.
This is expected because the net effect of the magnetic field is simply to shift the
threshold from $c_s^2$ to the one given by Eq.~(\ref{threshold-mag}) irrespective
of the shape of the probability distribution function.

From \ref{fig:B0_constraints_Gaussian-NG-compare}, 
for the non-Gaussian distribution function~\eqref{eq:rescaled_NG_PDF}, we estimate $B_p \lesssim 0.13 - 0.15 {\rm \mu G}$ on scale $k_p \sim 10^{17} {\rm Mpc}^{-1}$ and $B_p \lesssim 0.08 {\rm \mu G}$ on scale $k_p \sim 10^{8} - 10^{11} {\rm Mpc}^{-1}$. 
\ref{fig:B0_constraints_Gaussian-NG-compare} also compares our constraints (black and blue curves) with Saga et al.~\cite{2020-Saga.Hiroyuki.Yokoyama-JCAP} (shown in gray curves).
We would like to mention that our constraints incorporate new factors which
are not included in the previous work Ref.~\cite{2020-Saga.Hiroyuki.Yokoyama-JCAP}: 
(i) effect of the magnetic field in the threshold density contrast and (ii) variation of relativistic degrees of freedom.

Having derived the constraints from PBHs abundance, let us briefly discuss the necessary relations to incorporate other small-scale observational probes to constrain the PMFs.
For the constraints from GWB and magnetic reheating, we follow Refs.~\cite{2017-Saga.etal.Yokoyama-MNRAS,2018-Saga.etal-PRD}. However, as we have discussed in the previous section, concerning the variation of relativistic degrees of freedom with respect to temperature changes, the correct relation can be obtained by using Eq.\eqref{eq:rho_gamma-T-relation}. As we have shown, including the variation of the relativistic degrees of
freedom significantly affects the estimation of PBHs abundance. 
Since the stochastic GW background is also affected by the relativistic degrees of freedom, 
we must modify the relevant equation obtained in Refs.~\cite{2017-Saga.etal.Yokoyama-MNRAS,2018-Saga.etal-PRD} for the precise estimation of the upper limit on the amplitude of PMFs.
Within our framework, the energy density of GWB at the present epoch (corresponding to Ref.\cite{2018-Saga.etal-PRD}) is given as
\begin{align}\label{eq:modified_GWB-relation}
    \Omega_{\rm GW} (\eta_0, k) = \frac{ R_{\gamma} (\eta)^2}{512 \pi^2} \left( \frac{B^2}{\rho_{\gamma, 0}} \right)^2 \left( \frac{g_{*s}(T)}{g_{*s} (T_0)} \right)^{8/3} \left( \frac{k}{H_0} \right)^2 a_{\rm eq}^2 h_T^2 (k,\eta_{\rm eq} ) \left( \frac{k}{k_{\rm p}} \right)^2 \left( 1 +  \frac{k^2}{4 k_p^2} \right)^2 \Theta_H \left( 1 - \frac{k}{2 k_p} \right)
\end{align}
where $h_T (k,\eta_{\rm eq})$ is the transfer function of GWs given in Ref.~\cite{2018-Saga.etal-PRD}. Note that Eq.~\eqref{eq:modified_GWB-relation} tells us that given the upper bounds on $\Omega_{\rm GW}$, we can put an upper limit on the amplitude of PMFs.
\begin{figure}
\centering
\includegraphics[height=3.2in]{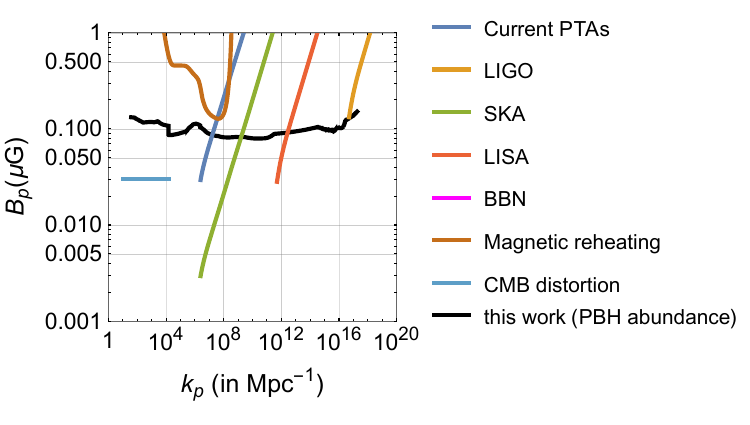} 
\caption{Upper limits on the primordial magnetic field $B_p$.}
\label{fig:all_constraints_plot}
\end{figure}
Following Ref.~\cite{2018-Saga.etal-PRD}, we focus on the direct detection measurements of GWB from pulsar timing arrays (PTAs) and SKA, space-based observatories like LISA, and ground-based observatories such as  LIGO, SKA, and LISA. Note that the PTA and LIGO bounds are obtained from the observed results; however, SKA and LISA are expected upper bounds in the future.

Another interesting effect of the PMFs on small scales is the magnetic reheating, recently proposed in Ref.~\cite{2017-Saga.etal.Yokoyama-MNRAS}. This mechanism is based on the fact that the dissipation of PMFs changes the baryon-to-photon number ratio and can constrain the PMFs on small scales. Again, for this case as well, within our framework, the injected energy fraction to the CMB energy by decaying PMFs should also be modified to 
\begin{align}\label{eq:modified_MR-relation}
    \frac{\Delta \rho_{\gamma} }{\rho_{\gamma} }  = \frac{ B^2 }{ 8 \pi \rho_{\gamma , 0} } \left( \frac{g_{*s}(T)}{g_{*s} (T_0)} \right)^{4/3}   \left[ e^{-\frac{2 k^2}{k_D^2 (z_i)}} - e^{-\frac{2 k^2}{k_D^2 (z_{\mu})}} \right]
\end{align}
where $z_{\mu} = 2\times 10^6$ and $z_{\nu} = 10^9$ are the redshifts corresponding to the CMB distortion era and neutrino decoupling era~\cite{2017-Saga.etal.Yokoyama-MNRAS}. The damping scale $k_D$ is related to the redshift as $k_D (z) = 7.44 \times 10^{-6} (1+z)^{3/2} \rm{Mpc^{-1}}$. 
\ref{fig:all_constraints_plot} shows the comparison of the constraints on the amplitude of the PMFs on small scales and comparison with various observations such as GW, magnetic reheating, CMB distortion, and BBN.
We would like to mention that the constraints in \ref{fig:all_constraints_plot} give a more precise estimation of the upper bounds on the amplitude of PMFs than presented in Ref.\cite{2020-Saga.Hiroyuki.Yokoyama-JCAP}. This is because one should consider two key issues: increase in the threshold density contrast due to PMFs and the variation of relativistic degrees of freedom with temperature.
\section{Conclusion and Discussion}
There are early-universe models which predict production of primordial magnetic fields on small scales. Those magnetic fields will dissipate in the early times and do not exist in the present universe, which makes probing such fields non-trivial.  In this work, we have systematically studied the procedure to constrain the amplitude of PMFs in the RD era from the abundance of PBHs. Given that anisotropic stress due to the PMFs sources the primordial curvature perturbations, for sufficiently strong PMFs the density perturbations might become large enough to gravitationally collapse to form PBHs in the RD era. However, we have shown that this picture is not completely correct because the magnetic field increases the threshold value of the density contrast required for the formation of PBHs. 
We computed this correction term to the threshold density contrast (due to magnetic field) and revised the constraints by using the new threshold.
Furthermore, if we naively extrapolate the new threshold to very small scales (for $k_p \gtrsim 10^{18}~{\rm Mpc}^{-1}$), the threshold density contrast becomes greater than one, 
which might stop the formation of PBHs. Since our estimation is based on the simple Newtonian analysis, more sophisticated works
based on general relativity are needed to reach a robust conclusion. Furthermore, GR corrections to the threshold 
which would modify both the first and the second terms on the right-hand side of Eq.(\ref{eq:deltaB-cs-rB})
by ${\cal O}(1)$ factors change significantly the PBH abundance,
which provides another motivation for studies based on GR.

To make the estimates more precise as compared to the previous studies, we have also considered the variation of relativistic degrees of freedom with respect to the formation epoch and shown that it affects the variance by $\mathcal{O}(2)$ on smaller scales. 
Thus, by taking into account the effect of magnetic fields on the threshold density contrast, variation of relativistic degrees of freedom, and using the most updated constraints on PBHs formation parameter, we estimated an upper-limit $0.13 - 0.15 {\rm \mu G}$ on scale $10^{17} {\rm Mpc}^{-1}$ and $ 0.08 {\rm \mu G}$ on scale $10^{8} - 10^{11} {\rm Mpc}^{-1}$ for non-Gaussian distribution function.
We also compared our constraints with those of other small-scale probes such as GWB, CMB distortion, magnetic reheating, and BBN. These constraints also help to support or rule-out certain classes of early-time magnetogenesis models.

%
\appendix

\section{Derivation of Eq.~(\ref{threshold-mag})}\label{app-rB-calc}
In this appendix, we provide derivation of the threshold (\ref{threshold-mag}).
Incorporating the effect of variation of relativistic DOF, ${\bar r_B}$ is given by
\begin{align}\label{eq-1}
& {\bar r_B} \equiv \frac{\rho_B}{\rho_{\rm tot}} = \frac{\rho_B}{\rho_{\gamma}} \frac{\rho_{\gamma}}{\rho_{\rm tot}}= \frac{B_f^2 }{8\pi \rho_{\gamma,0}} \left( \frac{a (T)}{a (T_0)} \right)^4  \left( \frac{g_{* s} (T) }{g_{*s} (T_0)} \right)^{4/3} R_{\gamma} (\eta) = \frac{B_p^2 }{8\pi \rho_{\gamma,0}} \left( \frac{g_{* s} (T) }{g_{*s} (T_0)} \right)^{4/3} R_{\gamma} (\eta)  .
\end{align}
\begin{figure}[h]
\centering
\includegraphics[height=2.8in]{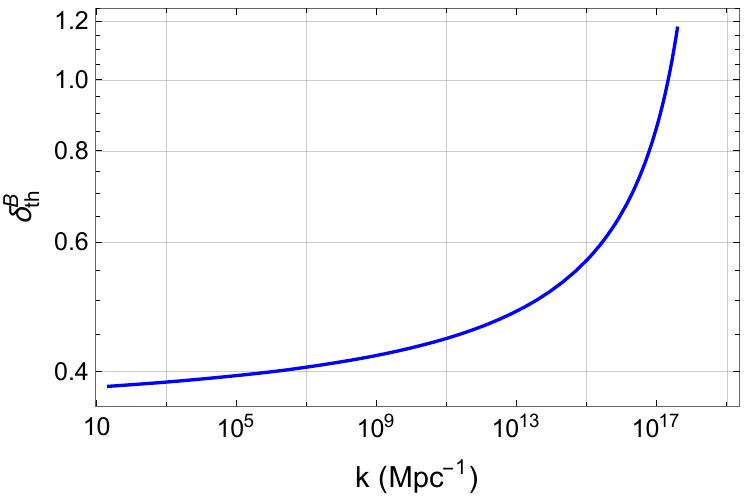} 
\caption{The behaviour of $\delta_{th}^B$.}
\label{fig:deltath-B}
\end{figure}
Then, using Eq.~(\ref{eq:sigmaB_xiR-simplified}) as $\sigma_B$,
we obtain
\begin{equation}
\label{appeq-rB}
\frac{\bar r_B}{\sigma_B} \approx 2.9\times \frac{1}{\xi (\eta)}.
\end{equation}
Substituting this result into Eq.~(\ref{eq:threshold}), we finally obtain 
the desired threshold
\begin{align}
\label{appeq-deltaB}
\delta_B > \frac{c_s^2  }{\left( 1-  \frac{5.8  }{  \xi(\eta) }  \right)} .
\end{align}
This is an important relation where RHS gives the threshold density contrast $\delta_{\rm th}^B$ required for the formation of PBHs due to magnetized overdense region. 
To see the effect of the correction term due to magnetic field, 
we plot Eq.~\eqref{appeq-deltaB} in Fig.~\ref{fig:deltath-B}, which shows that for $ k \gtrsim 10^{18} {\rm Mpc}^{-1}$ the threshold becomes greater than unity.

\section{Detailed analysis of the behaviour of quantity $\xi (\eta) R_{\gamma} (\eta)$}\label{appsec:xi-R_calculations}
\begin{figure}[h]
\centering
\includegraphics[height=2in]{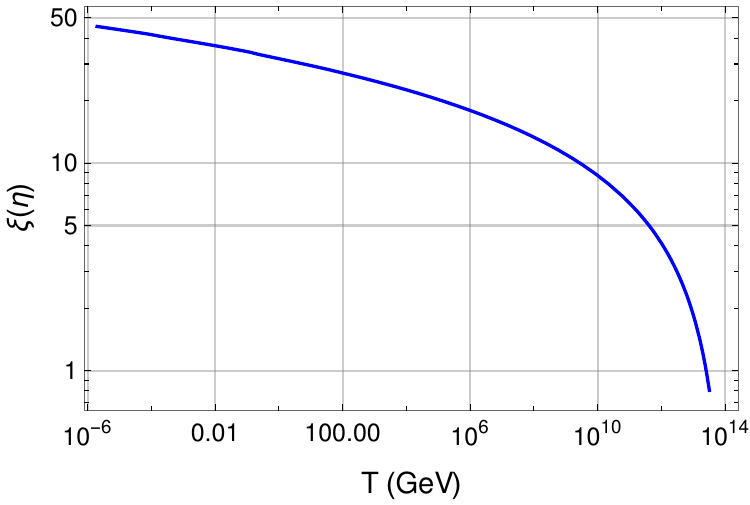} 
\centering
\includegraphics[height=2in]{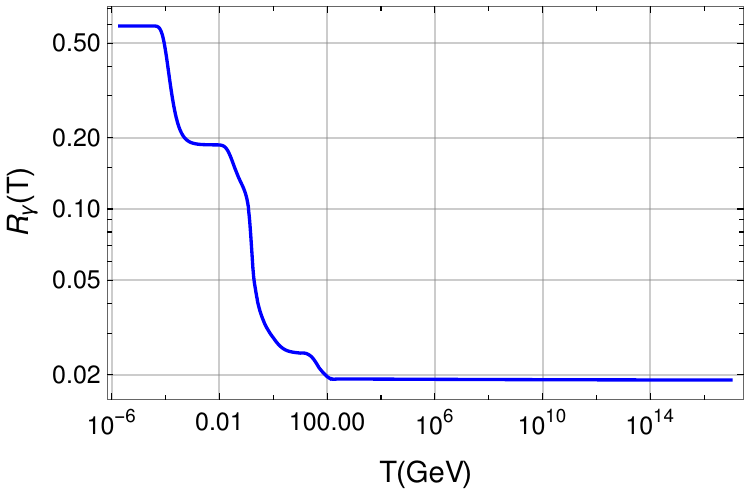} 
\caption{This figure (in LogLog-scale) shows the behaviour of $\xi (T)$ (left plot) and $R_{\gamma} (T)$ (right plot).}
\label{figapp:xi-Rgamma-T14}
\end{figure}

In this appendix, we provide a prescription to compute the quantity $\xi (\eta) R_{\gamma}(\eta)$ which appears in the expression of $\sigma_B^2$ in Eq.\eqref{eq:sigmaB_xiR-simplified}. We show that this quantity varies significantly for the scales of our interest and affect the formation of PBHs, hence as a consequence the constraints on the PMFs. 
Note that the estimation of $\xi (\eta)$ requires the precise values of conformal time $\eta$ at a given epoch and at the time of magnetic field generation ($\eta_B$). Therefore, we first derive the relation between $\eta$ and the temperature $(T)$ of the Universe, which allows us to compute the relevant quantities at any epoch (or scale) of interest.
Let us consider the following equations that describe the dynamics and thermal properties of the Universe~\cite{Book-Kolb.Turner,Book-Mukhanov,Book-Rubakov.Dmitry}, 
\begin{align}
\label{appeq:friedmanEq}
    H^2 &= \frac{8\pi G}{3} \rho \\
\label{appeq:thermoEq}
    \rho &= \frac{\pi^2}{30} g_* (T) T^4 \\
\label{appeq:entropyEq}
    g_{*s} (T) T^3 a^3 &= \rm{constant} 
\end{align}
where $g_* (T)$ and $g_{*s} (T)$ are relativistic degrees of freedom at temperature T for the radiation energy density $\rho (T)$ and entropy density $s(T)$, respectively. 
In this work, we follow the data provided by Ref.\cite{2018-Saikawa.Shirai-JCAP} for the relativistic degrees of freedom $g_{*} (T)$ and $g_{*s} (T)$.
In our estimates we use the following numerical values of the scale factor and wavenumber at matter-radiation equality i.e., $T_{\rm eq} \sim 1 {\rm eV}$:
\begin{align}
    a_{\rm eq} = 4.15\times 10^{-5} (\Omega_m h^2)^{-1}, \qquad k_{\rm eq} = 0.073 \Omega_m h^2 ~\rm{Mpc^{-1}}
\end{align}
 where $\Omega_m = 0.3106 \pm 0.011$ and $ h = 0.6770 \pm 0.0081$~\cite{Book-Dodelson-2003}, and in this work `eq' represents the quantities evaluated at matter-radiation equality.
 Using Eq.\eqref{appeq:entropyEq}, we obtain
\begin{align}\label{appeq:aT}
    a(T) = \left( \frac{g_{*s (T_{\rm eq})}}{g_{*s} (T)} \right)^{1/3} \left( \frac{T_{\rm eq}}{T} \right) a_{\rm eq}
    \implies  
    da(T) = -\left( \frac{g_{*s (T_{\rm eq})}}{g_{*s} (T)} \right)^{1/3} T_{\rm eq} a_{\rm eq} \left( 1 + \frac{T \frac{d g_{*s}(T)}{dT}}{g_{*s} (T)} \right) \frac{dT}{T^2}  .
\end{align}
The conformal time $\eta $ is defined as 
\begin{align}\label{appeq:eta-aprime}
    \eta \equiv \int^{t} \frac{dt^{\prime}}{a(t^{\prime})} = \int^{a} \frac{da^{\prime}}{{a^{\prime}}^2 H (a^{\prime})}
\end{align}
where $t$ is the cosmic time. Using Eqns.(\ref{appeq:friedmanEq},\ref{appeq:thermoEq},\ref{appeq:entropyEq},\ref{appeq:aT}) in Eq.\eqref{appeq:eta-aprime}, we obtain
\begin{align}\label{appeq:eta-T-relation}
    \eta (T) = \frac{1}{a_{\rm eq} T_{\rm eq} } \sqrt{\frac{90}{8\pi^3 G}} \int^{\infty}_{T} \frac{dT^{\prime}}{{T^{\prime}}^2} \left( \frac{g_{*s} (T^{\prime})}{g_{*s} (T_{\rm eq})} \right)^{1/3} \frac{1}{ \sqrt{g_* (T^{\prime})}} \left( 1 + \frac{T^{\prime} \frac{ dg_{*s} (T^{\prime}) }{dT^{\prime}} }{3 g_{*s} (T^{\prime})} \right) .
\end{align}
We would like to mention that Eq.\eqref{appeq:eta-T-relation} is a \emph{key equation} which gives the estimate of $\eta$ at a give $T$.
Following the above calculations and results, we can now estimate the functions $\xi (\eta (T))$ and $ R_{\gamma} (T)$, which are shown in Fig.~\ref{figapp:xi-Rgamma-T14}.
Since our fundamental variable is wavenumber $k$, it is more convenient
to estimate all the relevant quantities as a function of $k$ by relating the mode with $k$ 
with its horizon crossing time defined by $k=a(T) H(T)$.
Using Eqns.(\ref{appeq:friedmanEq}, \ref{appeq:entropyEq}, \ref{appeq:aT}), we obtain
\begin{align}\label{appeq:kp-T-relation}
    k = \sqrt{\frac{8\pi^3 G}{90}} \sqrt{g_* (T)} \left( \frac{g_{*s (T_{\rm eq})}}{g_{*s} (T)} \right)^{1/3}  T_{\rm eq} a_{\rm eq} T .
\end{align}
By eliminating $T$ from Eqns.(\ref{appeq:eta-T-relation}, \ref{appeq:kp-T-relation}),
we can relate $\eta$ with $k$.
Eq.(\ref{appeq:kp-T-relation}) allows us to determine $\xi (k)$ and $R_{\gamma} (k) = \frac{\rho_{\gamma}}{\rho_{\rm tot}} = 2/g_* (k)$.

\acknowledgments
AK thanks Avijit Chowdhury and Joseph P Johnson for discussions. AK would like to thank IIT Bombay for the financial support. AK would also like to acknowledge the financial support from the Indo-French Centre for the Promotion of Advanced Research (CEFIPRA) grant no. 6704-4 under the Collaborative Scientific Research Programme.
This work was supported by the JSPS KAKENHI Grant
Number JP23K03411 (TS). 
Finally, we would like to thank the referee for his or her insightful comments and suggestions that have helped to improve the manuscript.

\bibliographystyle{unsrt}
\bibliography{ReferencesJCAP}

\begin{thebibliography}{10}

\bibitem{1994-Kronberg-Rept.Prog.Phys.}
Philipp~P. Kronberg.
\newblock {Extragalactic magnetic fields}.
\newblock {\em Rept. Prog. Phys.}, 57:325--382, 1994.

\bibitem{1996-Beck.etal-ARAA}
Rainer Beck, Axel Brandenburg, David Moss, Anvar Shukurov, and Dmitry Sokoloff.
\newblock Galactic magnetism: Recent developments and perspectives.
\newblock {\em Annual Review of Astronomy and Astrophysics}, 34(1):155--206, 1996.

\bibitem{2002-Widrow-Rev.Mod.Phys.}
Lawrence~M. Widrow.
\newblock {Origin of galactic and extragalactic magnetic fields}.
\newblock {\em Rev. Mod. Phys.}, 74:775--823, 2002.

\bibitem{2004-Vallee-NAR}
Jacques~P. Vall\'ee.
\newblock {Cosmic magnetic fields \textendash{} as observed in the Universe, in galactic dynamos, and in the Milky Way}.
\newblock {\em New Astron. Rev.}, 48(10):763--841, 2004.

\bibitem{2004-Gaensler.Beck.etal-AstroRev}
Bryan~M Gaensler, R~Beck, and L~Feretti.
\newblock The origin and evolution of cosmic magnetism.
\newblock {\em New Astronomy Reviews}, 48(11-12):1003--1012, 2004.

\bibitem{2004-Giovannini-IJMPD}
Massimo Giovannini.
\newblock {The Magnetized universe}.
\newblock {\em Int. J. Mod. Phys. D}, 13:391--502, 2004.

\bibitem{2007-Barrow.etal-PhyRep}
John~D. Barrow, R.~Maartens, and Christos~G. Tsagas.
\newblock {Cosmology with inhomogeneous magnetic fields}.
\newblock {\em Phys. Rept.}, 449:131--171, 2007.

\bibitem{2011-Kandus.Kunze.Tsagas-PhysRept}
Alejandra Kandus, Kerstin~E. Kunze, and Christos~G. Tsagas.
\newblock {Primordial magnetogenesis}.
\newblock {\em Phys. Rept.}, 505:1--58, 2011.

\bibitem{2013-Durrer.Neronov-Arxiv}
Ruth Durrer and Andrii Neronov.
\newblock {Cosmological Magnetic Fields: Their Generation, Evolution and Observation}.
\newblock {\em Astron. Astrophys. Rev.}, 21:62, 2013.

\bibitem{2016-Subramanian-Arxiv}
Kandaswamy Subramanian.
\newblock {The origin, evolution and signatures of primordial magnetic fields}.
\newblock {\em Rept. Prog. Phys.}, 79(7):076901, 2016.

\bibitem{2010-Neronov.Vovk-Sci}
Andrii Neronov and Ievgen Vovk.
\newblock Evidence for strong extragalactic magnetic fields from fermi observations of tev blazars.
\newblock {\em Science}, 328(5974):73--75, 2010.

\bibitem{1988-Turner.Widrow-PRD}
Michael~S. Turner and Lawrence~M. Widrow.
\newblock {Inflation Produced, Large Scale Magnetic Fields}.
\newblock {\em Phys. Rev.}, D37:2743, 1988.

\bibitem{1991-Ratra-Apj.Lett}
Bharat Ratra.
\newblock {Cosmological 'seed' magnetic field from inflation}.
\newblock {\em Astrophys. J. Lett.}, 391:L1--L4, 1992.

\bibitem{1991-Vachaspati-PLB}
T.~Vachaspati.
\newblock {Magnetic fields from cosmological phase transitions}.
\newblock {\em Phys. Lett. B}, 265:258--261, 1991.

\bibitem{1993-Dolgov-PRD}
A.~D. Dolgov.
\newblock Breaking of conformal invariance and electromagnetic field generation in the universe.
\newblock {\em Phys. Rev. D}, 48:2499--2501, Sep 1993.

\bibitem{2007-Martin.Yokoyama-JCAP}
Jerome Martin and Jun'ichi Yokoyama.
\newblock {Generation of Large-Scale Magnetic Fields in Single-Field Inflation}.
\newblock {\em JCAP}, 01:025, 2008.

\bibitem{2016-Mukohyama-PRD}
Shinji Mukohyama.
\newblock Stealth magnetic field in de sitter spacetime.
\newblock {\em Phys. Rev. D}, 94:121302, Dec 2016.

\bibitem{2019-Kushwaha.Shankaranarayanan-PRD}
Ashu Kushwaha and S.~Shankaranarayanan.
\newblock {Galileon scalar electrodynamics}.
\newblock {\em Phys. Rev. D}, 101(6):065008, 2020.

\bibitem{2020-Bamba.Odintsov.etal-JCAP}
Kazuharu Bamba, E.~Elizalde, S.~D. Odintsov, and Tanmoy Paul.
\newblock {Inflationary magnetogenesis with reheating phase from higher curvature coupling}.
\newblock {\em JCAP}, 04:009, 2021.

\bibitem{2021-Giovannini-JCAP}
Massimo Giovannini.
\newblock {Palatini approach and large-scale magnetogenesis}.
\newblock {\em JCAP}, 11(11):058, 2021.

\bibitem{2021-Tripathy.etal-arXiv}
Sagarika Tripathy, Debika Chowdhury, Rajeev~Kumar Jain, and L.~Sriramkumar.
\newblock {On the challenges in the choice of the non-conformal coupling function in inflationary magnetogenesis}.
\newblock 11 2021.

\bibitem{2021-Nandi-JCAP}
Debottam Nandi.
\newblock {Inflationary magnetogenesis: solving the strong coupling and its non-Gaussian signatures}.
\newblock {\em JCAP}, 08:039, 2021.

\bibitem{Papanikolaou:2023nkx}
Theodoros Papanikolaou and Konstantinos~N. Gourgouliatos.
\newblock {Primordial magnetic field generation via primordial black hole disks}.
\newblock {\em Phys. Rev. D}, 107(10):103532, 2023.

\bibitem{Papanikolaou:2023cku}
Theodoros Papanikolaou and Konstantinos~N. Gourgouliatos.
\newblock {Constraining supermassive primordial black holes with magnetically induced gravitational waves}.
\newblock {\em Phys. Rev. D}, 108(6):063532, 2023.

\bibitem{Kamada:2020bmb}
Kohei Kamada, Fumio Uchida, and Jun'ichi Yokoyama.
\newblock {Baryon isocurvature constraints on the primordial hypermagnetic fields}.
\newblock {\em JCAP}, 04:034, 2021.

\bibitem{2012-Kawasaki.Kusakabe-PRD}
Masahiro Kawasaki and Motohiko Kusakabe.
\newblock Updated constraint on a primordial magnetic field during big bang nucleosynthesis and a formulation of field effects.
\newblock {\em Phys. Rev. D}, 86:063003, Sep 2012.

\bibitem{2019-Jedamzik.Saveliev-PRL}
Karsten Jedamzik and Andrey Saveliev.
\newblock Stringent limit on primordial magnetic fields from the cosmic microwave background radiation.
\newblock {\em Phys. Rev. Lett.}, 123:021301, Jul 2019.

\bibitem{2016-Kamada.Long-2}
Kohei Kamada and Andrew~J. Long.
\newblock {Baryogenesis from decaying magnetic helicity}.
\newblock {\em Phys. Rev. D}, 94(6):063501, 2016.

\bibitem{2016-Fujita.Kamada-PRD}
Tomohiro Fujita and Kohei Kamada.
\newblock {Large-scale magnetic fields can explain the baryon asymmetry of the Universe}.
\newblock {\em Phys. Rev. D}, 93(8):083520, 2016.

\bibitem{2021-Kushwaha.Shankaranarayanan-PRD}
Ashu Kushwaha and S.~Shankaranarayanan.
\newblock {Helical magnetic fields from Riemann coupling lead to baryogenesis}.
\newblock {\em Phys. Rev. D}, 104(6):063502, 2021.

\bibitem{2014-Kunze.Komatsu-JCAP}
Kerstin~E. Kunze and Eiichiro Komatsu.
\newblock {Constraining primordial magnetic fields with distortions of the black-body spectrum of the cosmic microwave background: pre- and post-decoupling contributions}.
\newblock {\em JCAP}, 01:009, 2014.

\bibitem{2017-Saga.etal.Yokoyama-MNRAS}
Shohei Saga, Hiroyuki Tashiro, and Shuichiro Yokoyama.
\newblock {Magnetic reheating}.
\newblock {\em Mon. Not. Roy. Astron. Soc.}, 474(1):L52--L55, 2018.

\bibitem{2018-Saga.etal-PRD}
Shohei Saga, Hiroyuki Tashiro, and Shuichiro Yokoyama.
\newblock {Limits on primordial magnetic fields from direct detection experiments of gravitational wave background}.
\newblock {\em Phys. Rev. D}, 98:083518, Oct 2018.

\bibitem{2020-Saga.Hiroyuki.Yokoyama-JCAP}
Shohei Saga, Hiroyuki Tashiro, and Shuichiro Yokoyama.
\newblock {Limits on primordial magnetic fields from primordial black hole abundance}.
\newblock {\em JCAP}, 05:039, 2020.

\bibitem{2002-Mack.Kahniashvili.etal-PRD}
Andrew Mack, Tina Kahniashvili, and Arthur Kosowsky.
\newblock {Microwave background signatures of a primordial stochastic magnetic field}.
\newblock {\em Phys. Rev. D}, 65:123004, 2002.

\bibitem{2012-Suyama.Yokoyama-PRD}
Teruaki Suyama and Jun'ichi Yokoyama.
\newblock {Metric perturbation from inflationary magnetic field and generic bound on inflation models}.
\newblock {\em Phys. Rev. D}, 86:023512, 2012.

\bibitem{2010-Shaw.Lewis-PRD}
J.~Richard Shaw and Antony Lewis.
\newblock {Massive Neutrinos and Magnetic Fields in the Early Universe}.
\newblock {\em Phys. Rev. D}, 81:043517, 2010.

\bibitem{Young:2014ana}
Sam Young, Christian~T. Byrnes, and Misao Sasaki.
\newblock {Calculating the mass fraction of primordial black holes}.
\newblock {\em JCAP}, 07:045, 2014.

\bibitem{2004-Green.etal-PRD}
Anne~M. Green, Andrew~R. Liddle, Karim~A. Malik, and Misao Sasaki.
\newblock {A New calculation of the mass fraction of primordial black holes}.
\newblock {\em Phys. Rev. D}, 70:041502, 2004.

\bibitem{2018-Sasaki.etal-CQG}
Misao Sasaki, Teruaki Suyama, Takahiro Tanaka, and Shuichiro Yokoyama.
\newblock {Primordial black holes\textemdash{}perspectives in gravitational wave astronomy}.
\newblock {\em Class. Quant. Grav.}, 35(6):063001, 2018.

\bibitem{2020-Carr.Florian-arXiv}
Bernard Carr and Florian Kuhnel.
\newblock {Primordial Black Holes as Dark Matter: Recent Developments}.
\newblock 6 2020.

\bibitem{2016-Nakama.Suyama-PRD}
Tomohiro Nakama and Teruaki Suyama.
\newblock {Primordial black holes as a novel probe of primordial gravitational waves. II: Detailed analysis}.
\newblock {\em Phys. Rev. D}, 94(4):043507, 2016.

\bibitem{1975-Carr-ApJ}
Bernard~J. Carr.
\newblock {The Primordial black hole mass spectrum}.
\newblock {\em Astrophys. J.}, 201:1--19, 1975.

\bibitem{Shibata:1999zs}
Masaru Shibata and Misao Sasaki.
\newblock {Black hole formation in the Friedmann universe: Formulation and computation in numerical relativity}.
\newblock {\em Phys. Rev. D}, 60:084002, 1999.

\bibitem{2005-Musco.Miller.Rezzolla-CQG}
Ilia Musco, John~C. Miller, and Luciano Rezzolla.
\newblock {Computations of primordial black hole formation}.
\newblock {\em Class. Quant. Grav.}, 22:1405--1424, 2005.

\bibitem{2013-Harada.etal-PRD}
Tomohiro Harada, Chul-Moon Yoo, and Kazunori Kohri.
\newblock {Threshold of primordial black hole formation}.
\newblock {\em Phys. Rev. D}, 88(8):084051, 2013.
\newblock [Erratum: Phys.Rev.D 89, 029903 (2014)].

\bibitem{Nakama:2013ica}
Tomohiro Nakama, Tomohiro Harada, A.~G. Polnarev, and Jun'ichi Yokoyama.
\newblock {Identifying the most crucial parameters of the initial curvature profile for primordial black hole formation}.
\newblock {\em JCAP}, 01:037, 2014.

\bibitem{Musco:2018rwt}
Ilia Musco.
\newblock {Threshold for primordial black holes: Dependence on the shape of the cosmological perturbations}.
\newblock {\em Phys. Rev. D}, 100(12):123524, 2019.

\bibitem{Young:2019osy}
Sam Young.
\newblock {The primordial black hole formation criterion re-examined: Parametrisation, timing and the choice of window function}.
\newblock {\em Int. J. Mod. Phys. D}, 29(02):2030002, 2019.

\bibitem{Escriva:2019nsa}
Albert Escriv\`a.
\newblock {Simulation of primordial black hole formation using pseudo-spectral methods}.
\newblock {\em Phys. Dark Univ.}, 27:100466, 2020.

\bibitem{1961-Chandrasekhar-Book}
Subrahmanyan {Chandrasekhar}.
\newblock {\em {Hydrodynamic and hydromagnetic stability}}.
\newblock 1961.

\bibitem{1966-Strittmatter-MNRAS}
P.~A. {Strittmatter}.
\newblock {A note on gravitational instability in a magnetic medium}.
\newblock {\em MNRAS}, 131:491, January 1966.

\bibitem{2019-He.Suyama-PRD}
Minxi He and Teruaki Suyama.
\newblock {Formation threshold of rotating primordial black holes}.
\newblock {\em Phys. Rev. D}, 100(6):063520, 2019.

\bibitem{2012-Byrnes.etal-JCAP}
Christian~T. Byrnes, Lukas Hollenstein, Rajeev~Kumar Jain, and Federico~R. Urban.
\newblock {Resonant magnetic fields from inflation}.
\newblock {\em JCAP}, 03:009, 2012.

\bibitem{2020-Patel.etal-JCAP}
Teerthal Patel, Hiroyuki Tashiro, and Yuko Urakawa.
\newblock {Resonant magnetogenesis from axions}.
\newblock {\em JCAP}, 01:043, 2020.

\bibitem{2022-Sasaki.Vardanyan.etal-arXiv}
Misao Sasaki, Valeri Vardanyan, and Vicharit Yingcharoenrat.
\newblock {Super-horizon resonant magnetogenesis during inflation}.
\newblock {\em Phys. Rev. D}, 107(8):083517, 2023.

\bibitem{2018-Saikawa.Shirai-JCAP}
Ken'ichi Saikawa and Satoshi Shirai.
\newblock {Primordial gravitational waves, precisely: The role of thermodynamics in the Standard Model}.
\newblock {\em JCAP}, 05:035, 2018.

\bibitem{2018-Inomata.etal-PRD}
Keisuke Inomata, Masahiro Kawasaki, Kyohei Mukaida, and Tsutomu~T. Yanagida.
\newblock {Double inflation as a single origin of primordial black holes for all dark matter and LIGO observations}.
\newblock {\em Phys. Rev. D}, 97(4):043514, 2018.

\bibitem{2023-LISACosmologyWorkingGroup}
Eleni Bagui et~al.
\newblock {Primordial black holes and their gravitational-wave signatures}.
\newblock 10 2023.

\bibitem{2021-Carr.etal-ReptProgPhys}
Bernard Carr, Kazunori Kohri, Yuuiti Sendouda, and Jun'ichi Yokoyama.
\newblock {Constraints on primordial black holes}.
\newblock {\em Rept. Prog. Phys.}, 84(11):116902, 2021.

\bibitem{Book-Kolb.Turner}
Edward~W. {Kolb} and Michael~S. {Turner}.
\newblock {\em {The early universe}}, volume~69.
\newblock 1990.

\bibitem{Book-Mukhanov}
Viatcheslav Mukhanov.
\newblock {\em Physical Foundations of Cosmology}.
\newblock Cambridge University Press, 2005.

\bibitem{Book-Rubakov.Dmitry}
Valery~A. Rubakov and Dmitry~S. Gorbunov.
\newblock {\em {Introduction to the Theory of the Early Universe}: {Hot big bang theory}}.
\newblock World Scientific, Singapore, 2017.

\bibitem{Book-Dodelson-2003}
Scott Dodelson.
\newblock {\em {Modern Cosmology}}.
\newblock Academic Press, Amsterdam, 2003.

\end{thebibliography}

\end{document}